\documentclass[reqno,superscriptaddress,nofootinbib,preprint, onecolumn]{revtex4}
\usepackage[centertags]{amsmath}
\usepackage{amsfonts}
\usepackage{amssymb}
\usepackage{amsthm}
\usepackage{newlfont}
\usepackage{mathrsfs}
\usepackage{mathtools}
\usepackage{euscript}
\usepackage{graphicx}
\usepackage{enumerate}
\usepackage{todonotes}
\usepackage{color}
\usepackage{floatrow}
\usepackage{tikz}
\usepackage{pgf}
\usepackage{wrapfig}
\usepackage{subfigure}
\usepackage{amscd}
\usepackage{hyperref}
\usepackage{changes}

\newcommand{\ie}{\textit{i.e.}}

\def\bea{\begin{eqnarray}}
\def\eea{\end{eqnarray}}
\def\ba{\begin{array}}
\def\ea{\end{array}}

\def\lsim{\,\lower2truept\hbox{${< \atop\hbox{\raise4truept\hbox{$\sim$}}}$}\,}
\def\gsim{\,\lower2truept\hbox{${> \atop\hbox{\raise4truept\hbox{$\sim$}}}$}\,}
\newcommand{\id}{\textrm{d}}

\begin{document}

\title{Active velocity processes with \\suprathermal stationary distributions and long-time tails} 

\begin{abstract}
	When a particle moves through a spatially-random force field, its momentum may change at a rate which grows with its speed.  Suppose moreover that a thermal bath  provides friction which gets weaker for large speeds, enabling high-energy localization.  The result is a unifying framework for the emergence of heavy tails in the velocity distribution, relevant for understanding the power-law decay in the electron velocity distribution of space plasma or more generally for explaining non-Maxwellian behavior of driven gases.  We also find long-time tails in the velocity autocorrelation, indicating persistence at large speeds for a wide range of parameters and implying superdiffusion of the position variable.   
\end{abstract}
\author{Tirthankar Banerjee}
\email{tirthankar.banerjee@kuleuven.be}
\affiliation{Instituut voor Theoretische Fysica, KU Leuven, Belgium}
\author{Urna Basu}
\affiliation{Raman Research Institute, Bengaluru, India}
\author{Christian Maes}
\affiliation{Instituut voor Theoretische Fysica, KU Leuven, Belgium}

\maketitle
\section{Introduction}
A particle moving in weak contact with a thermal bath experiences friction and noise in an equilibrated fashion as expressed in the fluctuation-dissipation relation~\cite{fdt,kubo-fdt,chandler,Balki}. Brownian motion is the standard example, obtaining a steady fluctuating motion as described in a Langevin dynamics where possibly other conservative forces are added.  The steady state evolution is then said to run under the condition of detailed balance~\cite{lewis}.  In some physically interesting cases however the particle may {\em also} be subject to an external nonconservative force field.  Such a field can be the coarse-grained result of underlying more complicated processes, such as arising from a turbulent environment or from the influence of biologically-active matter.  The force may be averaging out to zero either in space or in time and yet has an influence on the particle motion.  There is no extra systematic force, no drift is added, but the environment provides extra sources of uncompensated noise.  That noise need not be Gaussian in general and can be interpreted as excess dynamical activity transmitted from the environment to the particle.  We refer to such environments (possibly massless) as {\it active media}.  The active velocity processes in the title are the inertial motions of particles in such active media.  We come back to the relation with (models for bio-)active particles in Sec. \ref{compac}. Conceptually, our models are for example closer to those in ~\cite{gel1,gel2} where the dynamics is studied of a tracer particle in an active gel.  The main question of the present paper is to investigate the resulting steady velocity distribution of the tagged particle and its relaxation properties.

To be clear about the physical situation, the tagged particle (probe) is part of a dilute bath to which we can associate a temperature, and the active medium is external and to be modelled as an extra random force field.  Let us discuss these two ingredients and our assumptions on them first separately.

{\bf The heat bath}: we suppose a dilute bath of particles where the friction is essentially determined from the two-body scattering cross section.  Therefore, the way the scattering depends on the particle kinetic energy is an essential input.  We assume that the scattering gets vanishingly small at high energy, which is a condition of (high) energy localization. That happens in many cases of interest, e.g. in the regime of Coulomb scattering~\cite{bian-2014}: charged particles at high (kinetic) energy tend to keep their energy when moving fast.

{\bf The random force field}: we imagine a spatial distribution of a nonequilibrium forcing.  The latter may be caused by moving optical~\cite{opt-ref}, acoustic~\cite{ac-twzr} or mechanical walls or by spacetime-dependent external force fields more generally.  An important assumption is to take the force field spatially-mixing of  spatial average equal to zero and having a finite correlation length.  Such a condition can be called spatially chaotic. At the high speeds that we will consider, we ignore the time-dependence of the force field.

Probes moving in such an active medium evolve under an inertial dynamics for which the fluctuation-dissipation relation is violated, allowing net energy transfer from the medium to the probe which may then be dissipated in the thermal bath.  Active velocity processes have appeared before in models of velocity resetting, e.g. as first considered for Fermi acceleration~\cite{Fermi-prl-1949}, or in depot and Rayleigh-Helmholtz models~\cite{Schweitzer-prl-98,Roman-epj-2012,lin,roman}
or in models of Taylor dispersion \cite{cvdb} and Ulam ping pongs \cite{Ulam}.  
We can think of tagged grains in agitated matter or of electrons in a driven plasma.
Each time, the tagged particle (probe), while itself passive, is weakly coupled to a thermal bath in the presence of a nonequilibrium forcing. 

The central result of the present work is a unifying framework for suprathermal tails in the velocity distribution and (nonequilibrium) long-time tails. The exponents follow from the nature of the activity and from dependence of the scattering cross section on the kinetic energy. As we show, the high-energy localization combined with the chaotic fluctuating force field is responsible for interesting nontrivial behavior that is seen in nature, relevant for astrophysical plasmas~\cite{lin-2003} and in excited granular media~\cite{Losert-chaos-99,menon-prl-2000,olafsen-pre-2002}, or in general for the dynamical properties of tagged particles in a thermal bath under spatially-mixing or chaotic external driving conditions~\cite{lander}. 

The essential mathematics is contained in the setup of Sec. \ref{gens}. We compare that setup with (bio-)active particle models in Sec. \ref{compac}. Section \ref{stoac} introduces the key-step. The influence of the external nonconservative force field in the high-speed regime effectively results in a weak coupling limit with a nonequilibrium bath.  The resulting noise is then Gaussian alright but no friction equilibrates it.  The phenomenon is known as stochastic acceleration \cite{stoacs1,stoacs2,stoacs3}: when the particle speed is sufficiently large,  the change in its momentum over even a small time-interval fluctuates around zero following the central limit theorem. It induces an extra diffusion in velocity space with an amplitude $\propto 1/v$ decaying with the particle speed.  In the end, the result of that analyis gives a three-dimensional Fokker-Planck description in which the effects of stochastic acceleration are combined with (high-energy) localization: for a dilute gas or plasma, the probability density $\mu(v,t)$ for the speed  satisfies
\bea\label{fknon}
\frac{\partial \mu}{\partial t}(v,t)&=&
\frac 1{v^2}\frac{\partial}{\partial v}\left(\gamma(v)\,v^2\,\left[v\,\mu(v,t) + \frac{k_BT}{m}\frac{\partial \mu}{\partial v}(v,t)\right]\right) \cr
&&+   \frac 1{v^2}\frac{\partial}{\partial v}\left(v^2\,\frac{A^2\,L}{2v\,m^2}\frac{\partial \mu}{\partial v}(v,t)\right) 
\eea
where $m$ is the mass of the particle and $\gamma(v)$ its friction coefficient for moving through the bath at temperature $T$.  The random force field is felt by the last term, where $A$ is its amplitude and $L$ its correlation length.  For example and to be detailed below starting with Sec. \ref{sut}, from \eqref{fknon} the power-law tail in the stationary velocity distribution is easily derived when $\gamma(v) \propto v^{-3}$ for large speeds $v$ such as obtained from the Rutherford formula for Coulomb scattering.

\section{General setup} \label{gens}
We consider a dilute medium of particles with mass $m$ where the interaction is described on the one-particle level in terms of a friction $\gamma$ and a white noise at temperature $T.$  We consider the velocity distribution in terms of a density $\rho(\mathbf v, t)$ with respect to the volume element $\id^3 \mathbf v$.  
In the absence of any external force, assuming that the medium is spatially homogeneous and that the initial density $\rho(\mathbf v, t=0)$ only depends on the speed $v$, $\rho(\mathbf v,t)= \rho(v, t)$ evolves in time $t$ according to the 3-dimensional Fokker-Planck equation
\begin{equation}\label{FPeq}
\frac{\partial \rho}{\partial t}(v,t) =
\frac 1{v^2}\frac{\partial}{\partial v}\left(v^2\,\gamma(v)\,\left[v\,\rho(v,t) + \frac{k_BT}{m}\frac{\partial \rho}{\partial v}(v,t)\right]\right)
\end{equation}
It is clear from \eqref{FPeq} that because of the imposed Einstein relation between the friction $m\gamma(v)$ and the noise variance $mk_BT \gamma(v)$, the stationary density for \eqref{FPeq} is Gaussian $\rho(v) \propto \exp [- mv^2/2k_BT]$ for arbitrary $\gamma(v)>0$.  We emphasize that this scenario is valid as well for a dilute plasma where the particles (ions, electrons,...) mutually interact with Coulomb forces and the friction behaves as $\gamma(v) \propto v^{-3}, v\uparrow \infty$ following the Rutherford scattering formula where the cross-section $\gamma(v) / v \propto (v^2)^{-2}$ is inversely proportional to the square of the energy.  More generally, in the present paper we take
\begin{equation}\label{g}
\gamma(v) = \gamma_0 \,\left[1 + \left(\frac{v}{v_R}\right)^\delta\,\right]^{-1}
\end{equation}
parameterized by the linear friction constant $\gamma_0>0$ and where $v_R$ is a reference speed beyond which the friction starts to decrease. The important parameter giving the decay \eqref{g} with the speed is $\delta$.  For $\delta>0$,  the scattering cross section for the particle in the thermal environment decreases like $v^{1+\delta}$ for large $v$.  Coulomb scattering gives $\delta = 3$ but we expect that depending on the material and shape of the particles in inelastic short-range scattering, values with $\delta<3$ become available.  At any event, when $\delta> 1$, high-energy localization takes place as then the strength $\gamma(v)\,v$ of the friction force  decays as $K^{(1-\delta)/2}$ when the kinetic energy $K\propto v^2$ grows large.  Nevertheless, for every $\delta$ the stationary distribution for \eqref{FPeq} is Maxwellian.

To represent the active medium we add a force field $\mathbf f(\mathbf r,t), \mathbf r
\in \mathbb{R}^3$. The evolution is then governed by the Langevin equation
\begin{eqnarray}
\dot{\mathbf r}_t &=& \mathbf v_t\nonumber\\
m\dot{\mathbf v}_t &=& -m\gamma(v_t)\, \mathbf v_t + \mathbf f(\mathbf r_t,t) + k_B T\, \gamma'(v_t)\,
\mathbf e_t + \sqrt{2m\gamma(v_t) k_BT}\,\boldsymbol{\xi}_t \label{rtv3}
\end{eqnarray}
In the last term lives the standard white noise $\boldsymbol{\xi}_t$.
The third term on the right-hand side of \eqref{rtv3} involving $\gamma' = \frac{d\gamma}{dv}$ arises from choosing the It\^o-convention and $\mathbf e_t = \mathbf v_t/v_t$ is the unit vector in the direction of the velocity. That term would vanish when writing (\ref{rtv3}) in the Stratonovich sense~\cite{itovstrat}.  In that way, when $\mathbf f\equiv 0$ (passive case), the Fokker-Planck equation corresponding to \eqref{rtv3}  reduces to \eqref{FPeq}. \\

\begin{figure}[t]
 \centering
 \includegraphics[width=8 cm]{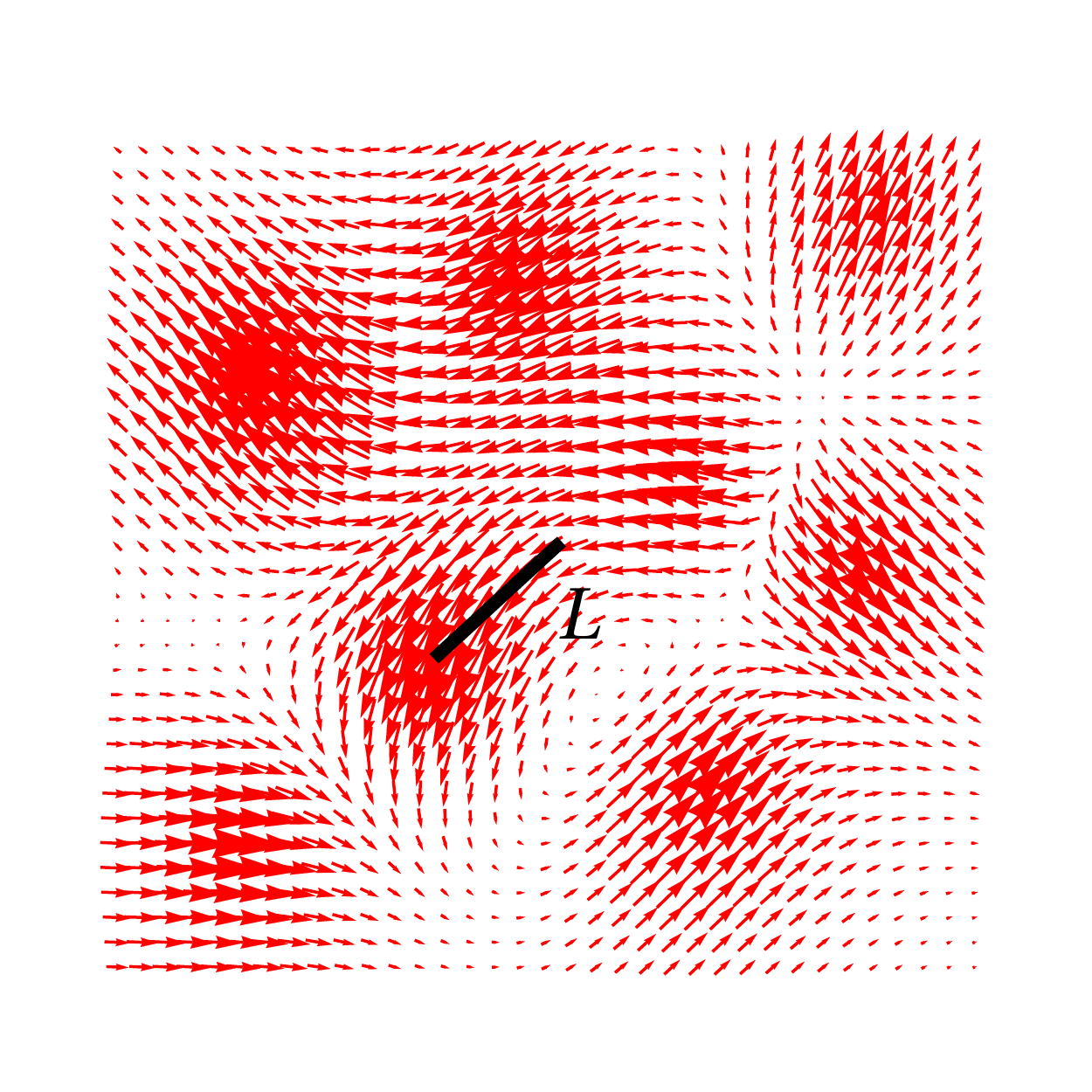}
 \caption{Discrete realization of a random force field $\mathbf f(\mathbf r,t)$ which auto-decorrelates over distances of order $L$. The tagged particle moves through it and at high speeds experiences a diffusive acceleration.}  
 \label{fig:force_field}
\end{figure}

The force field $\mathbf f(\mathbf r,t)$ is fundamentally arising from Newtonian forces, e.g. for electrons as a Lorentz force from a time-dependent electromagnetic field or  for granular particles as collisions with a vibrating wall, but we think of it as sufficiently chaotic to motivate its randomness.  We treat it as a quenched random field, homogeneous in spacetime and spatially isotropic.  To make it locally constant over a spatial range $L$ and with a persistence time $\nu_0^{-1}$, the statistics of the field $\mathbf f=(f_i)$ is modeled with
\bea
\langle \mathbf f(\mathbf r,t)\rangle &=& 0, \cr
\langle f_i(\mathbf r,t)\,f_j(\mathbf r',t')\rangle &=& 
C\left(\frac{\mathbf r-\mathbf r'}{L}, \nu_0\,(t-t')\right)\delta_{ij}~~ \label{ffo}
\eea 
for Kronecker-delta $\delta_{ij}$ and with a function $C$ showing exponential decay in space with range $L$, and changing in time at rate $\nu_0$. The length $L$ and rate $\nu_0$ indicate the space and time-scales over which its direction changes; see Fig.~\ref{fig:force_field} for a schematic representation of $\mathbf f(\mathbf r,t)$. The dynamics \eqref{rtv3} together with \eqref{ffo}
 specify mathematically what we mean by the active velocity processes mentioned in the title of the paper. 
For our purposes we only need the spatial decay and the intrinsic temporal dependence can be ignored. In \eqref{ffo} we have only one spatial scale $L$ but the arguments below hold more generally for multi-scaled fields, as long as they are sufficiently mixing to apply the central limit theorem (next).  In any event, the force $\mathbf f$ in \eqref{rtv3} is a second (non-thermal) source of noise in the particle dynamics. Correspondingly, there is a nonequilibrium steady condition for the dynamics \eqref{rtv3} with a stationary velocity distribution $\rho(v)$ that we investigate next for its tails at large speed $v$ and for its relaxational behavior in Sec. \ref{stime}.  We continue its analysis in Sec. \ref{stoac}.

\section{Connection to active particle models}\label{compac}

The dynamics \eqref{rtv3} is essentially different from active particle models for self-propelled motion in biology \cite{marchetti,ramaswamy,bechinger-review}. An important difference is that active particles in a biological context have an overdamped dynamics. Moreover, our analysis essentially uses the dependence of the friction $\gamma(v)$ on the speed, which in the overdamped limit would imply a dependence of the mobility on the speed, which is certainly not the main ingredient for e.g. bacterial motion. In biology the particle itself is deemed active because of a persistent speed  where the direction of the velocity is subject to a colored or discrete noise. 

We use run-and-tumble models  as an inspiration and example for a random force field in Sec. \ref{tum}.  Tumbling, and run-and-tumble models,  ~\cite{thompson-lattice, berg-2004} have been considered before to model active particles such as via self-propulsion in bacteria or in nanomotors.  In the present paper we have no internal nonequilibrium degrees of freedom coupled to translation, but we use tumbling as one way to model the activity of the external medium: the run-and-tumble in Sec. \ref{tum} concerns the incurred force. It is exciting to find tumbling forces relevant in understanding physical phenomena beyond the usually studied biological applications.
Apart from inspiration, there is also significance of our work to active matter when we think of the random force field as {\em created} by the presence of bio-active particles or active tissue in which we immerse a passive (underdamped) probe.  We already mentioned the examples of motion in an active gel, \cite{gel1,gel2}.  Then, the emergence of suprathermal distributions and long-time tails give new signatures of activity.

\section{Stochastic acceleration}\label{stoac}
To investigate the consequences of the force field \eqref{ffo}, we zoom in on the effect of $\mathbf f$ on the change in momentum of a moving particle.  Because of $\mathbf f$, the tagged particle following \eqref{rtv3} will, at time $t,$ incur a(n additional) change of momentum
\begin{equation}\label{del}
\mathbf \Delta_\epsilon(t) = \int_t^{t+\epsilon} \mathbf f(\mathbf  r_t + \mathbf v_t\, s,s) \,\id s 
\end{equation}
over each small enough time-interval $\epsilon$ (so that the velocity of the particle is not changing considerably). We fix that arbitrarily small $\epsilon$ for the rest of the argument. 
As the random field is homogeneous and isotropic, the distribution of the change $\mathbf \Delta_\epsilon(t)$ is independent of $\mathbf r_t$ and, with unit vector $\mathbf e_t$ in the direction of the velocity, we have
\begin{equation}\label{indis}
\int_t^{t+\epsilon} \mathbf f(\mathbf  r_t + \mathbf v_t\, s,s) \,\id s  \mathop{=}_{}^{\cal D} \frac 1{v_t}\,\int_0^{v_t\epsilon} \mathbf f(x\,\mathbf e_t,x/v_t) \,\id x 
\end{equation}
with the equality meant in the sense of probability distribution.  Hence, by taking $v_t$ large compared to $L/\epsilon$, \eqref{del} integrates to zero with a correction following the central limit theorem.  From \eqref{indis}, that means that \eqref{del} is of order $\sqrt{\epsilon}$ for large enough $v_t$ with, from \eqref{ffo}, a finite variance which is proportional to $L/v_t$.
%
As a conclusion, from the assumed chaoticity \eqref{ffo} of the external medium we infer that for large speeds $v_t$ there is a constant $A\neq 0$ depending on the function $C$ and possibly on the dimension, such that for \eqref{del},
\begin{equation}\label{sa}
\mathbf \Delta_\epsilon(t)  \mathop{=}_{}^{\cal D} A\sqrt{\frac{\epsilon L}{v_t}}\, \mathbf Z
\end{equation}
where $\mathbf Z$ is a 3-dimensional standard normal random variable.  Since that argument can be repeated for arbitrarily small $\epsilon$, we have obtained what is needed for the velocity process to become indistinguishable from a Markov diffusion when running at sufficiently high speeds. It replaces the force $\mathbf f$ in \eqref{rtv3} by a white noise.  The equality \eqref{sa} is meant in the sense of distributions and is derived in Sec. \ref{tum} for a precise realization of forcing. Mathematically rigorous work for the more general case can be found in \cite{stoacs1,stoacs2,stoacs3}.  The main physical mechanism goes back to the phenomenon of  Taylor dispersion~\cite{Taylor-53, Taylor-54a, Taylor-54b,cvdb}, from where the general concept of stochastic or turbulent acceleration arises~\cite{soaf,pope-2002,kimura-2010,nonM}.   

As a result, for large speeds $v_t$ we effectively have two white noises, the thermal noise from \eqref{rtv3} and the stochastic acceleration from \eqref{sa}. The corresponding differential equation for the probability density (denoted by $\mu(v,t)$ to make a difference with the dynamics \eqref{rtv3}) exactly becomes \eqref{fknon}.
Note that the calculation \eqref{del}--\eqref{sa} of the stochastic acceleration followed the It\^o-sense, estimating ${\mathbf v}_{t+\epsilon}- {\mathbf v}_t$, and hence we have no additional correction to the drift.\\
The stationary density $\mu(v)$ for \eqref{fknon} can be solved exactly,
\begin{equation}\label{sd2}
\mu(v) \propto \exp \left[- m\int^{v}_0 \id u\, \,\frac{u}{ k_BT+ A^2\,L/(2\gamma(u)\,u)}\right]
\end{equation}
The argument above can  be concluded by the statement that for large $v$, the stationary distribution $\rho(v)$ for the (original) dynamics \eqref{rtv3} equals the one from \eqref{fknon}--\eqref{sd2}, \ie, $\rho(v) \simeq \mu(v)$.  

\begin{figure*}[t] 
 \includegraphics[width=5.8 cm]{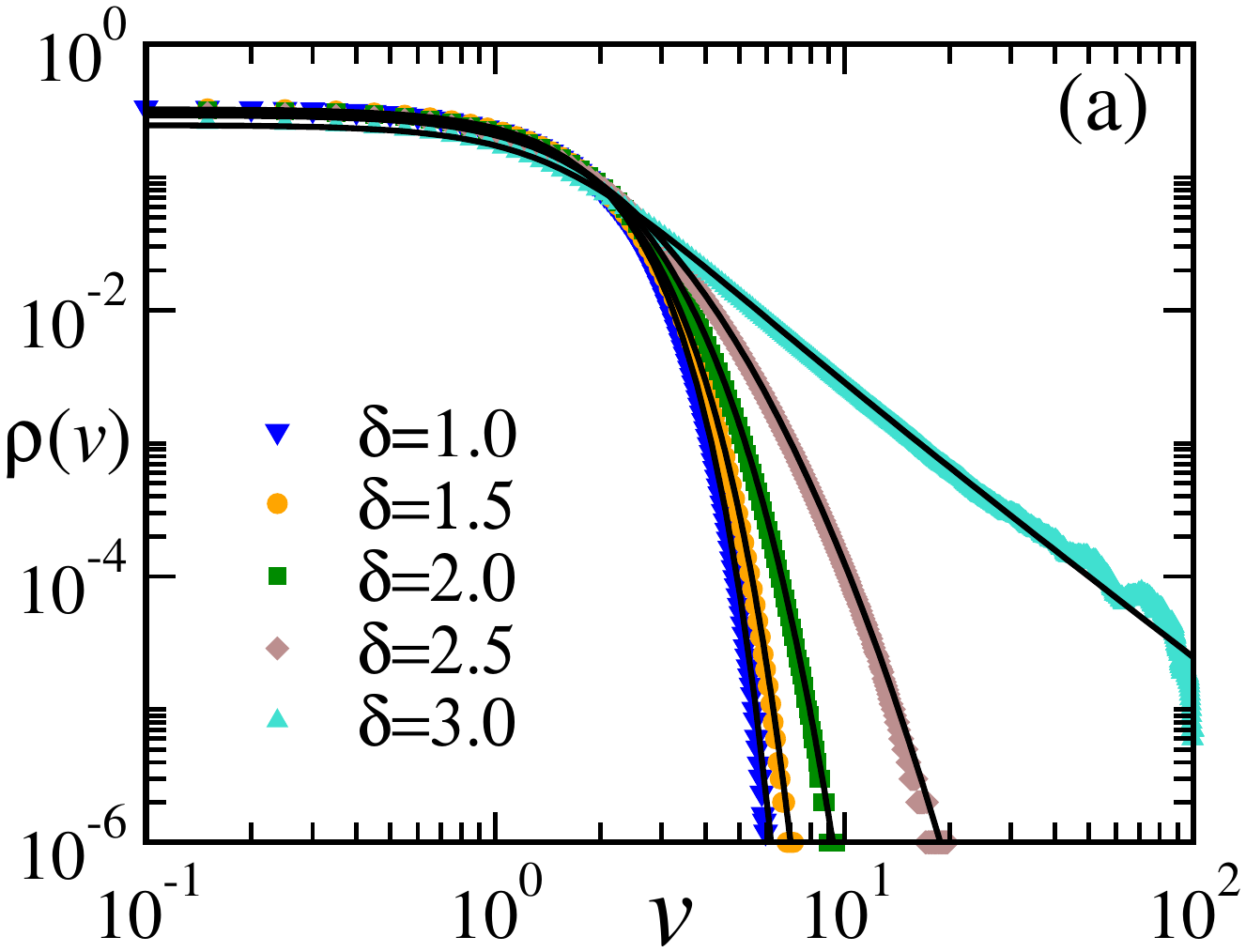}\includegraphics[width=11 cm]{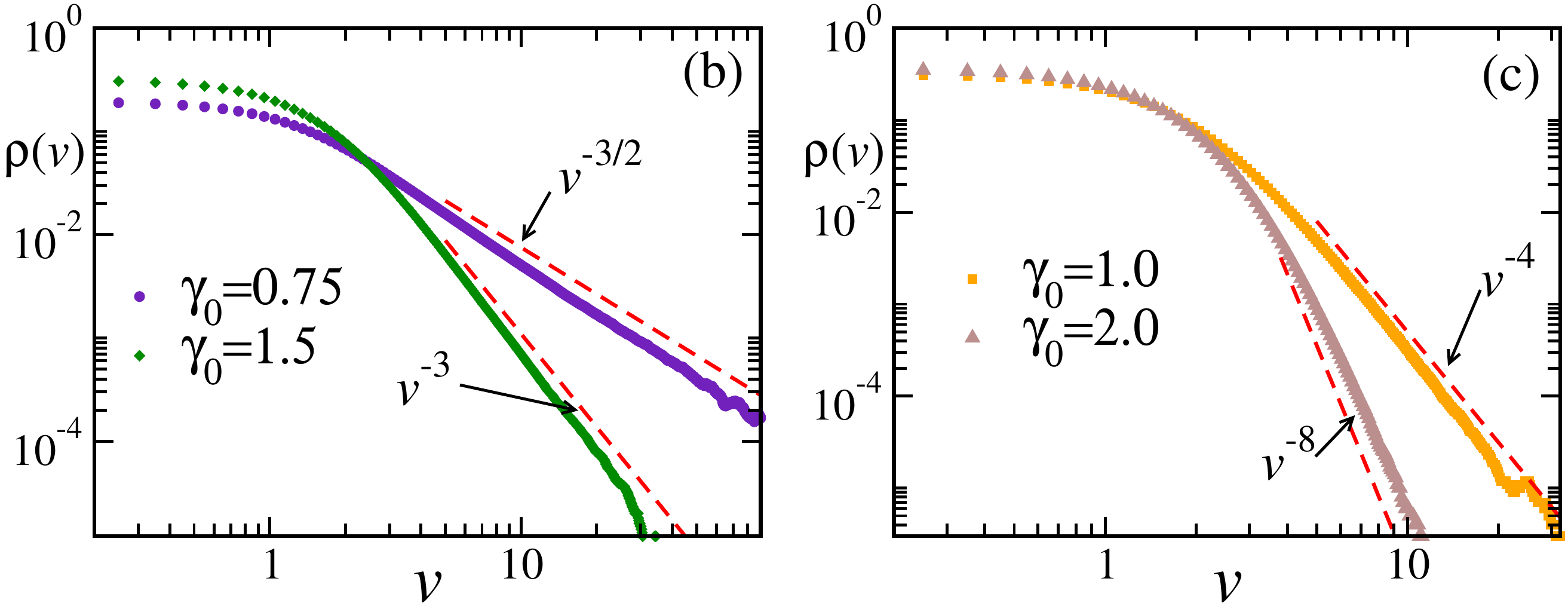} 	
\caption{(a) Plot of $\rho(v)$ {\it vs} $v$ with fixed $\gamma_0=1=L=A=T=\nu_0$ for different values of $\delta;$ see \eqref{ag}.  Note the transition from a Maxwellian (for $\delta=1.0$) to power law decay ($\delta=3.0$) via compressed ($\delta= 1.5$), simple ($\delta=2.0$) and stretched ($\delta=2.5$) exponential regimes for increasing $\delta$. Symbols represent data obtained from Monte-Carlo simulations while solid black lines correspond to the $\mu(v)$ obtained from evaluating (\ref{sd2}) numerically. (b) and (c) Plot of $\rho(v)$ {\it vs} $v$ for $\delta=3, A=1=T=\nu_0$. The power-law decay in the stationary velocity distributions is shown for  (b) $L=1$, and (c) $L=1/2.$ Various values of $\gamma_0$ following \eqref{ag} are plotted. The symbols correspond to the data obtained from numerical simulations and the red dashed lines indicate the theoretically predicted algebraic decay.}\label{figsR}
 \end{figure*}

\section{Suprathermal tails}\label{sut}

To understand the asymptotic behaviour of the stationary distribution $\rho(v)$ we use the explicit form \eqref{sd2} along with the friction $\gamma(v)$
in \eqref{g}. It turns out that the behaviour is qualitatively different for $\delta=3$ and $\delta<3$ while for $\delta > 3$ no stationary distribution exists for \eqref{fknon}.

Algebraic decay clearly appears from (\ref{sd2}) when $\delta=3.$ More specifically, when  $\delta =3$ in \eqref{g}, then
\begin{equation}\label{kap}
\rho(v) \sim v^{-2\kappa}, \qquad \kappa = m\, \frac{\gamma_0\,v_R^3}{L A^2}
\end{equation}
for  $v \gg v_R$ and $mv^2/2 \gg \kappa\,k_BT$. 
As a reminder, $\rho(v)$ must be multiplied with $4\pi v^2$ (from $\id^3 \mathbf v = 4\pi v^2\id v$), to get the normalized speed distribution. To have a finite variance (sometimes referred to as kinetic temperature) we thus need that $\kappa> 5/2$ which is consistent with \cite{lin-2003, lazar-2017}.  Observe also that the $\kappa$ in \eqref{kap} is a ratio of friction-parameters over activity-parameters: larger friction increases $\kappa$ while larger activity and persistence reduce $\kappa$.  In solar plasma, the reported values for $\kappa$ are around 5, while the onset of the power law happens at energies $ \simeq 0.1$ keV~\cite{battaglia-2015}. The term in the numerator $\gamma_0 v_R^3$ is essentially known from the Rutherford formula (or the mean-free length in dilute collisionless plasma of density about $10^6 m^{-3}$) to be about $0.14 \times 10^{-19}$ kg/s. Therefore our formula \eqref{kap} will be useful to estimate nonequilibrium aspects.  From the previously mentioned numerical values we get $LA^2 
\simeq 1.8 \times 10^{-49}$ $\text {kg}^2/\text {s}$, characterizing the effective driving force field in solar plasmas.\\
Suprathermal velocity distributions, where the high-energy tail is overpopulated with respect to the corresponding Maxwellian, have been observed in space plasmas~\cite{parker-pr-58,lin-1971,lin-2003} and there go under the name of kappa-distributions~\cite{kap,bian-2014,lazar-2017,lazar-2015}. The fact that an effective
diffusivity that depends inversely on the speed can produce suprathermal velocity distribution functions was already discussed,  e.g. in~\cite{bian-2014,nonM}, in the context
of highly-energetic space plasmas. A general formulation based on a Fokker-Planck equation was e.g. already given in \cite{parker-pr-58} but without obtaining the kappa-distribution \eqref{kap}.

Continuing with \eqref{sd2}, we predict a pure exponential decay for $\delta=2$, compressed exponential for $1<\delta<2$, stretched exponential for $2 < \delta < 3$ and Gaussian for $\delta \leq 1$.  In general, when  $1<\delta<3$ in \eqref{g}, then
\begin{equation}\label{stret}
\rho(v) \sim \exp \left[ - \frac{\kappa}{b}\left(\frac{v}{v_R}\right)^{2b}\right], \qquad b = \frac{3-\delta}{2}
\end{equation} 
again for large $v$.  When $\delta=1$ we recover the Maxwellian (Gaussian) behavior of \eqref{maxw} for large $v$ but with effective temperature $T + m\,v_R^2/(2k_B\kappa)$. 
The literature is vast and various modeling schemes and approximations have been offered. As an example we refer to the experimental results \cite{Losert-chaos-99,menon-prl-2000,olafsen-pre-2002} in excited granular media.\\

From the standpoint of statistical physics, the emergence of suprathermal tails due to a chaotic external force-field is new and unifies various phenomena.  We will next take an explicit example to illustrate the above scenario and to discuss long-time tails caused by emerging persistence of high speeds.

\section{Tumbling forces}\label{tum}

So far we have considered general active velocity processes where the incurred force on a tagged particle changes at a rate proportional to its speed.  We can imagine that along its trajectory there is a time-dependent force with local peristence time $L/v_t$ when the speed gets big.  In the rest of the paper we simplify that idea even further by taking a class of dynamics where the external force is randomly `tumbling'.  Such processes with tumbling forces provide an interesting illustration of the general setup and conclusions of the previous section.  By the greater simplicty of telegraphic noise~\cite{dkmps,weiss} a more detailed analysis becomes available while preserving the main physical idea.  In particular we predict a strong steady temporal autocorrelation, thus realizing long-time tails in a nonequilibrium environment.\\

To start with and for simplicity of notation we restrict ourselves to one spatial dimension.
The tumbling-force model in one dimension for a particle of mass $m=1$ with velocity $v_t \in \mathbb{R}$ at time $t$ is given by the Langevin equation (from now on, $k_B=1$)
\begin{eqnarray}\label{rtv}
\dot{x}_t &=& v_t\nonumber\\
\dot{v}_t &=& -\gamma(v_t)\, v_t + A\,\sigma_t + T\, \gamma'(v_t) + \sqrt{2\gamma(v_t) T}\,\xi_t\,\;
\end{eqnarray}
where $\xi_t$ is standard white noise in the It\^o-convention.  The external force has amplitude $A\geq 0$ and the tumbler $\sigma_t = \pm 1$ is taken to flip at a rate 
\begin{equation}\label{ag}
\alpha(v) = \nu_0 + L^{-1} \,|v|
\end{equation}
The flipping rate or the frequency that the incurred force changes direction is thus linearly increasing with its speed, consistent with then physical scenario of Sec. \ref{gens}  As before, the particle undergoes energy and momentum exchanges with a thermal bath at temperature $T\geq 0$ and nonlinear friction coefficient $\gamma(v) = \gamma(|v|) > 0$ given in \eqref{g}.

Mathematically, the dynamics \eqref{rtv} defines a Markov process $(v_t,\sigma_t)$ in velocity and tumble variables.    The joint probability on velocity $v_t \in \mathbb{R}$ and force $\sigma_t=\pm 1$ has a density $\rho_{\pm}(v,t)$ for time $t$. The corresponding differential equation for the probability density is 
\begin{eqnarray}\label{fpa}
\frac{\partial \rho_{\pm}}{\partial t} (v,t) &=& \frac{\partial}{\partial v}[\left(\gamma(v)\,v\, \mp A  - \gamma'(v)\,T\right)\rho_{\pm}(v,t)] + \alpha(v)\,[\rho_{\mp}(v,t) -\rho_{\pm}(v,t)]\cr
&& + ~ T\,\frac{\partial^2}{\partial v^2}\left(\gamma(v)\rho_{\pm}(v,t)\right), \quad v\in \mathbb{R}
\end{eqnarray}
Observe that for $A=0$ (passive case)  the Maxwellian
\begin{equation}\label{maxw}
\rho_{\pm}^{A=0}(v) \propto \exp [- v^2/2T]
\end{equation}
is the stationary (equilibrium) density, independent of the friction $\gamma(v)$.  For $A\neq 0$ there is a higher-order equation for $\rho(v,t) =\rho_{+}(v,t) + \rho_{-}(v,t)$ that determines the stationary velocity distribution $\rho(v)$ ($=\rho(v,t\rightarrow \infty)$).
We want to understand its behavior as $|v|\rightarrow \infty$ when $A\neq 0$, and how it depends on the friction $\gamma(v)$. The physical input that determines the interesting choices for $\alpha(v), \gamma(v)$ are in \eqref{g} and \eqref{ag}.  In what follows we often choose $\nu_0=1$ in \eqref{ag} setting a time-scale.


  The main idea to get a theoretical prediction for large $|v|$ is to follow Sec. \ref{stoac}  and to exploit that $\alpha(v)$ grows with $|v|$.  When $|v| \gg L\nu_0$, we may expect (extra) diffusive behavior induced by the activity.   Consider therefore the contribution of the tumbling force only, as in the updating
\begin{equation}\label{intins}
v_{t+\epsilon} = v_t + A \, \int_t^{t+\epsilon} \id s \,\sigma_s
\end{equation}
for fixed small $\epsilon$.
Note that the tumbling correlations are given by
$\langle \sigma_u\sigma_s\rangle =  e^{-2\alpha|u-s|}$
where we were allowed to take $\alpha=\alpha(v_t)$ constant for $0\leq u,s\leq \epsilon$ as $\epsilon$ is taken very small.  Therefore we have the variance $\left < (v_{t+\epsilon}-v_t)^2\right > = A^2\,\epsilon/\alpha.$

\begin{figure}[t]
	\centering
	\includegraphics[width=11 cm]{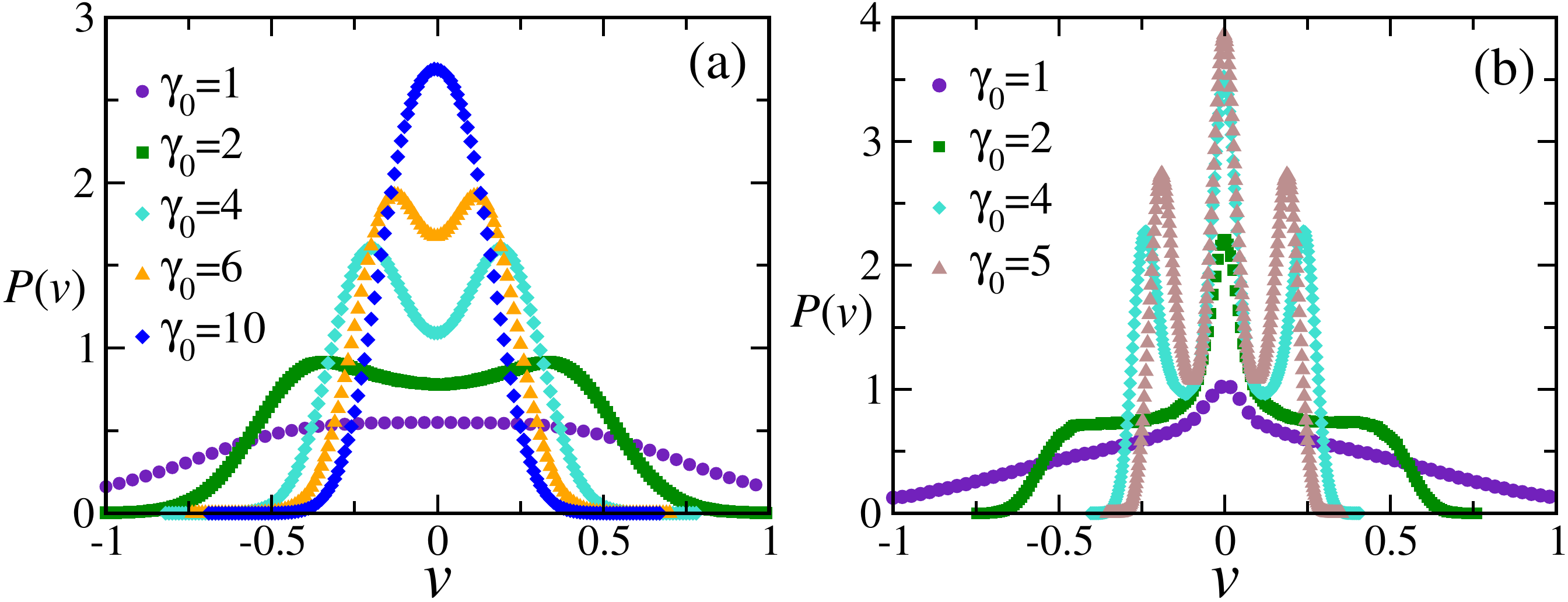}
	\caption{Behaviour of $\rho(v)$ for small $v$: (a) The bimodal stationary density appears low temperature (here $T=0.01$)for different values of $\gamma_0$ with $A=1=L$ and $\delta=3$. (b) Plot of $\rho(v)$ {\it vs.} $|v|$ for the three-state run-and-tumble process with a flipping rate $\alpha(v)/2$ between any two states. Trimodality appears for small $T$ (here $T=0.001$), with $A=1=L, \delta=3$.}
	\label{fig:tri}
\end{figure}

Moreover, in distribution,
\begin{eqnarray}\label{clt0}
 \int_t^{t+\epsilon} \id s \,\sigma_s~ && \mathop{=}_{}^{\cal D} \frac 1{\alpha}\,\int_0^{\alpha\epsilon} \id u \,\tilde{\sigma}_u\nonumber\\
 &&= \sqrt{\frac{\epsilon}{\alpha}}\,\frac 1{\sqrt{\alpha\epsilon}}\,\int_0^{\alpha\epsilon} \id u \,\tilde{\sigma}_u
 \end{eqnarray}
 where the process $\tilde{\sigma}_u$ runs with flip rate equal to one.  Hence, whenever $\alpha(v_t)\,\epsilon\gg 1$ we can apply the central limit theorem to $\frac 1{\sqrt{\alpha\epsilon}}\,\int_0^{\alpha\epsilon} \id u \,\tilde{\sigma}_u$ and continue from \eqref{intins} to get
\begin{equation}\label{clt}
v_{t+\epsilon} \simeq  v_t + A\,\sqrt{\frac{\epsilon}{\alpha(v_t)}}\,Z
\end{equation}
where $Z$ is a standard normal random variable. That is the stochastic acceleration \eqref{sa} as $\alpha(v) \sim L^{-1} |v|$ for large $|v|$.  Thus for large flipping rate $\alpha$, the tumble force can effectively be modeled by a white noise of strength $A^2/\alpha$; see also~\cite{dkmps}.  In the large-speed regime, the dynamics \eqref{rtv} appears thus replaceable by a passive Langevin dynamics $v_t$ (in It\^o-sense),
\begin{eqnarray}\label{eff-diff}
\dot{v}_t &=& -\gamma(v_t)\,v_t +  T\,\gamma'(v_t) + \sqrt{2\gamma(v_t) T}\;\xi_t^{(1)} +  \sqrt{\frac{A^2}{\alpha(v_t)}}\;\xi_t^{(2)}
\end{eqnarray}
 with two independent white noises $\xi_t^{(1)}$ and $\xi_t^{(2)}$ of zero mean and unit variance.  We repeat that the approximation (\eqref{eff-diff} instead of \eqref{rtv}) requires large $|v|$ (as we assume in \eqref{ag} that $\alpha(v)$ grows with $|v|$) and gets better for high enough $T$ to exclude the zero-$T$ cut-off $|\dot v|\leq \gamma_0|v| + A$; see below for more on that around Figs.~\ref{fig:tri}.
    
  \begin{figure}[t]
  	\centering
  	\includegraphics[width=11 cm]{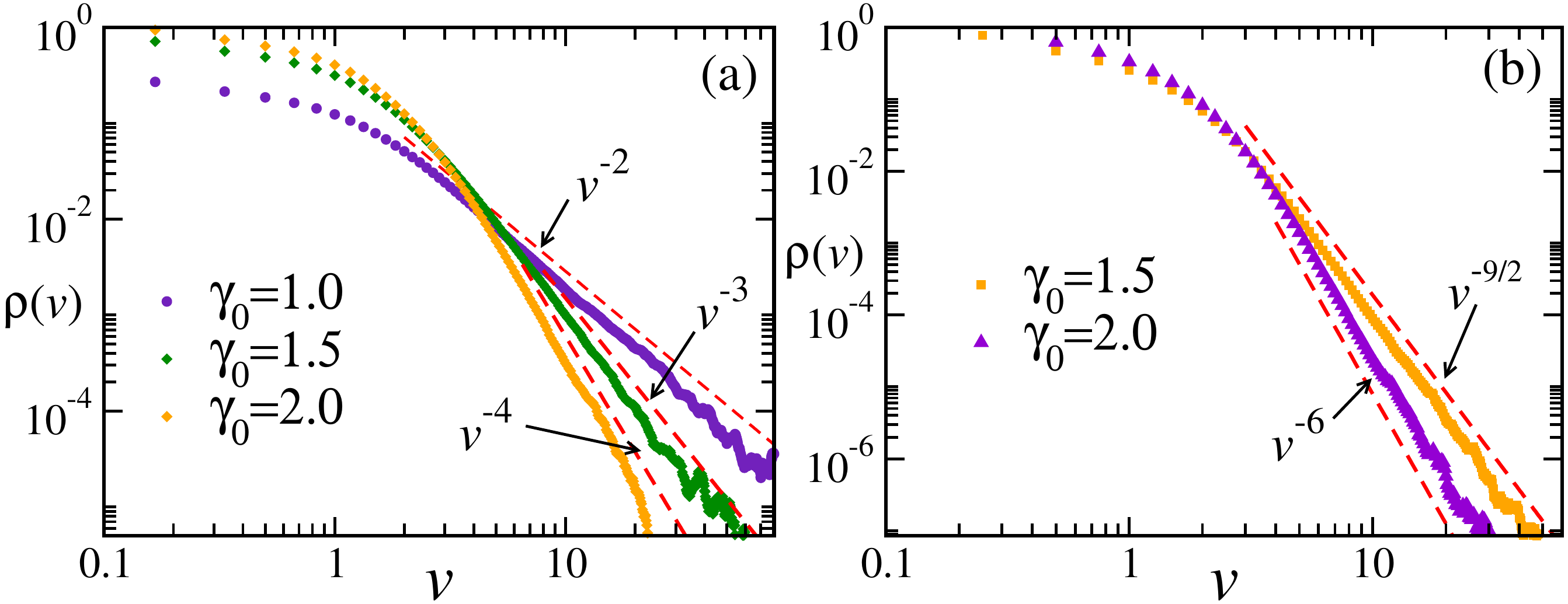}
  	\caption{Plot of $\rho(v)$ {\it vs} $v$  for varying $\gamma_0$ with $\delta=3$, $A=1=T$ (a) for $d=2$ and $\nu_0=1=L$, (b) for $d=3$ with $\nu_0=3/2$ and $L=2/3.$ Symbols show the  data from numerical simulations and the red dashed lines indicate corresponding analytical predictions.}\label{kap-2d}
  \end{figure}
 
In order to verify these predictions we have simulated the dynamics (\ref{rtv}) using the Euler-discretization scheme,
\begin{equation}\label{discr-dyn}
v_{t+\epsilon} = v_t -\epsilon [\gamma(v_t) v_t -  A \sigma_t -  T \gamma^{\prime}(v_t)] + \sqrt{2 \epsilon T \gamma(v_t)} \, Z
\end{equation}
where $Z$ is a random number drawn from the standard normal distribution. We use  $\epsilon=0.001$ for the time-step. In all the simulations following \eqref{discr-dyn}, $v$ is in units of $v_R$ which we put equal to $1$. We have also set $m=1=k_B$. The distributions are then obtained by averaging over at least $10^9$ samples in the steady state.
\par
In Fig.~\ref{figsR}(a), we plot $\rho(v)$ $vs$ $v$ for varying $\delta$ and compare the simulation results with a numerical evaluation of Eq.~\ref{sd2}. Fig.~\ref{figsR} (b) and (c) show plots of $\rho(v)$ $vs$ $v$ for varying $\kappa$ (fixed $\delta=3$). Note the excellent match between our analytical predictions and the corresponding simulation results.

\vspace{0.5cm}
Next we consider some generalizations of this simple model. First we can look at dimensions $d=2,3$. The dynamics is given by \eqref{rtv3} in the form,
  \[
  \dot{\mathbf v}_t = -\gamma(v_t)\, \mathbf v_t +  A\, \hat{\mathbf f}(t) + T\, \gamma'(v_t)\,
  \mathbf e_t + \sqrt{2\gamma(v_t) T}\,\boldsymbol{\xi}_t
  \]
  where $ \hat{\mathbf f}(t)$ is a Markov process taking values in the space of unit-vectors (points on the circle in $d=2$, or points on the unit sphere for $d=3$).  Uniformly at rate $R=2\alpha(v_t)/d$, a random unit vector is chosen.
Then, similar to \eqref{ffo}, $\langle  \hat {f}_i(u)\,  \hat f_j(s) \rangle = \frac{1}{d}\,\exp(-R\,|u-s|)\,\delta_{ij}, d=2,3$.  As before in \eqref{sa},  and in \eqref{clt} we have $\langle ({\mathbf v}_{t+\epsilon}-{\mathbf v}_t)^2 \rangle = \frac{A^2 \epsilon}{\alpha}$ and the $\kappa$ of \eqref{kap} is unchanged. The comparison with numerical results are presented in Fig.~\ref{kap-2d} (a) for $d=2$ and in Fig.~\ref{kap-2d}(b) for $d=3$.

  
  A second generalization from the case \eqref{rtv} is to allow  more than two values for $\sigma_t$. We skip the detailed calculations but
clearly all arguments are robust with respect to such changes. The suprathermal nature of the stationary velocity distribution is not affected, and we again get an algebraic decay of the velocity distribution $\rho(v)$ (not shown here).

We conclude that tumbling forces model the dynamics of particles in random force fields to produce heavy velocity tails. The flipping of the direction of the external force is easily imagined for Fermi--Ulam ping pong~\cite{Fermi-prl-1949,Ulam} or even in the case of granular gases under nonequilibrium driving.\\ 

As a final remark, it is also interesting to inspect where the tumbling fingerprint lies for small $|v|$ in the stationary distribution $\rho(v)$ of \eqref{rtv}. For low enough $T$, bimodality appears in the steady state distribution of $v$; see Fig.~\ref{fig:tri}(a). Note that at $T=0$, the particle resembles in velocity space a run-and-tumble particle in a harmonic trap which shows bimodality in its stationary behavior~\cite{bechinger-review, dkmps}. For small enough $T \neq 0$, this feature survives. For a fixed low $T$, however, $\rho(v)$ undergoes a shape transition from being highly localized near $v=0$ to a delocalized distribution as $\gamma_0$ is decreased from very large to small values. A large friction in effect makes the particle immobile. As $T$ is increased the thermal noise takes over and  the diffusive behavior leads to a broadening of the peaks, which eventually disappear for large enough temperatures. Similarly, when the tumbling variable  takes three values  $0,1,-1$, we obtain a trimodal distribution for small $v$ at sufficiently small $T$ for varying $\gamma_0$; see Fig.~\ref{fig:tri}(b). These features resemble well the results found in \cite{gel1,gel2}.

\begin{figure}[t]
 \centering
 \includegraphics[width=8 cm]{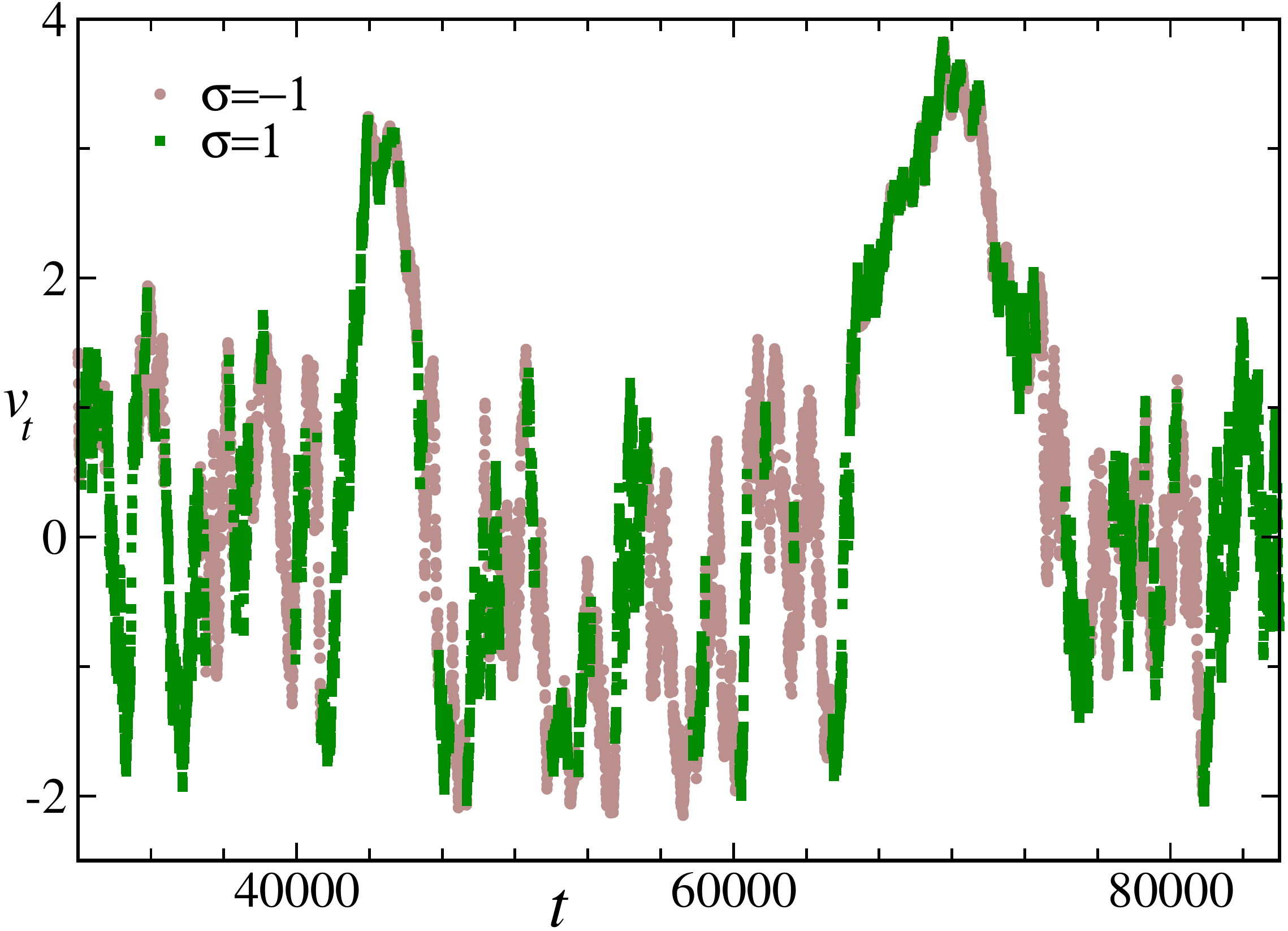}
 \caption{Time--evolution of the velocity $v_t$ along a typical trajectory of \eqref{rtv} in the steady state for $A=1=L=\nu_0$, $\delta=3, \gamma_0=2$. The light grey discs and the dark green squares correspond to $\sigma=-1$ and $1$, respectively. Switching between the two symbols represents tumblings of the active force.
 }
 \label{time-series}
\end{figure}

\section{Steady time-autocorrelation} \label{stime}
One may wonder whether the heavy tails in the velocity distribution are accompanied by long-time tails in the steady velocity autocorrelation,
\begin{equation}\label{t-corr-defn}
c(t) = \langle v_0\,v_t\rangle - \langle v_t\rangle\langle v_0\rangle
\end{equation}
We continue in one dimension. We consider the averaging being carried out in the steady state, so that $\langle v_t \rangle = 0 = \langle v_0 \rangle $.  To estimate the time-dependence of \eqref{t-corr-defn} we imagine drawing an initial velocity $v_0$ from $\rho(v) (\simeq \mu(v)$ under $|v| \gg L\nu_0$) in \eqref{sd2} and the question is to see at what time $v_t$ decorrelates wth $v_0$.  If $|v_0| \ll v_R$ (small initial speed), the friction induces a time-scale $\gamma_0^{-1}$ with exponentially fast decorrelation.  On the other hand, for large speeds $|v_0|$,  the friction is mostly absent and decorrelation happens after another time-scale. 
For the heuristics we refer to Fig.\ref{time-series} to observe a persistence in (large) speed.   We get a quantitative prediction by reconsidering (\ref{rtv}) for cases when friction and thermal effects are negligible and where the updating is given by (\ref{intins}).  Clearly, for no matter what $v_0>0$, when at time $t$,
\begin{equation}
\label{z}
\int_0^{t} \id s \,\sigma_s \in \left[- \frac{\Delta}{A}v_0\,,\, \frac{\Delta}{A}\,v_0 \right]
\end{equation}
then, $v_0v_t \simeq (1\pm \Delta)\,v_0^2$. where $\Delta \simeq 1/2$ is a dimensionless tolerance. 
Invoking the central limit theorem as in (\ref{clt}),
we are thus asked to estimate the probability that $\sqrt{\frac{t}{\alpha}}\,Z \in \left[- \frac{v_0}{2\,A}\,,\, \frac{v_0}{2\,A} \right]$, which amounts to evaluating the error-function at a value proportional to $\sqrt{\alpha/t}\;v_0/A = t^{-1/2}\;v_0^{3/2}/(\sqrt{L}A)$. We conclude that the event \eqref{z} occurs with high probability if $t \ll v_0^3/(L\,A^2)$.   Therefore,
 we predict that the time autocorrelation behaves as 
\bea
c(t) \simeq \int_{a\,(L A^2\,t)^{1/3}}^\infty\,\id v\, \rho(v)\,v^2 \label{eq:ct_int}
\eea
for some $a>0$, when $\rho(v)\,v^2 < 1/v$ decays sufficiently fast.  All other contributions decay faster in time.

\begin{figure}[t]
 \centering
 \includegraphics[width=11 cm]{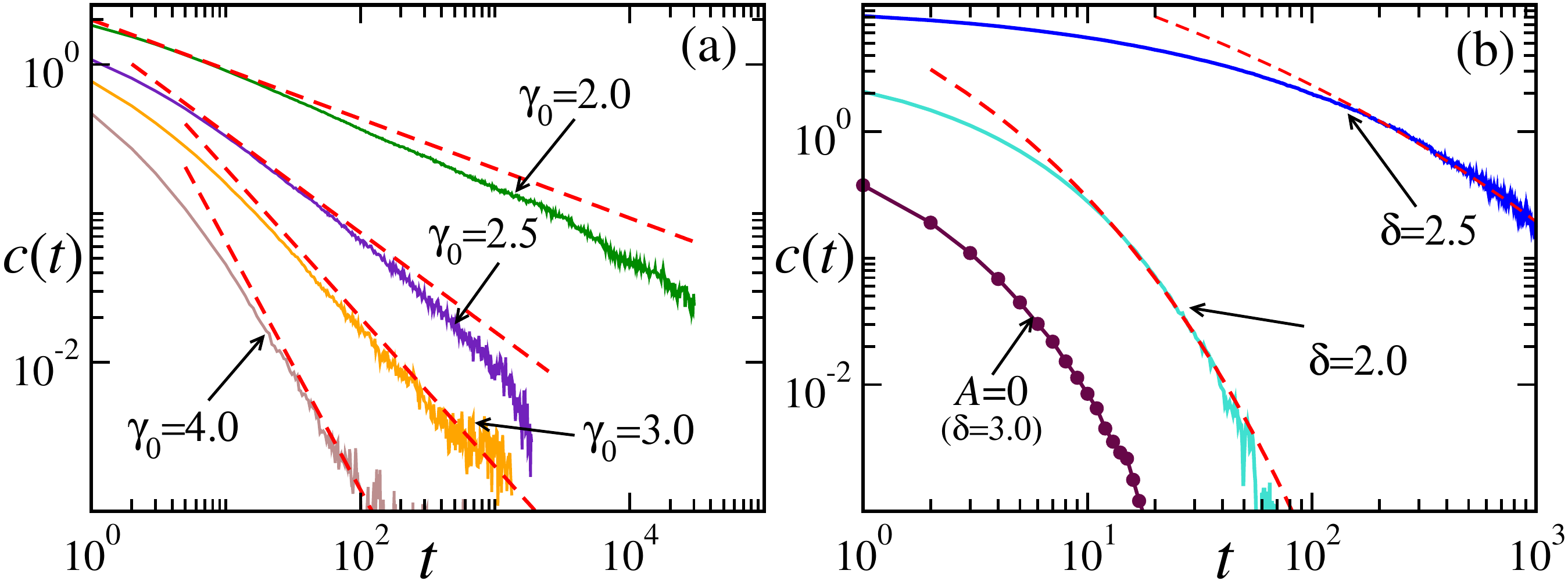}
 \caption{(a) Plot of $c(t)$ {\it vs} $t$ obtained from numerical simulations for $\delta=3$ and varying $\gamma_0$ with $L=1=A=T$. The red dashed lines indicate the analytical prediction (\ref{st-corr}). (b)  $c(t)$ {\it vs} $t$  for two other values of $\delta$; the red dashed lines indicate the best fit according to the prediction \eqref{strettime} where $\bar k$ has been used as a fitting parameter. Here $L=1=T$ and $\gamma_0=2=A$. The lowest curve corresponds to the equilibrium case $A=0$ for $\delta=3$ with $L=1=T=\nu_0$ and $\gamma_0=2.$}
 \label{ltt}
\end{figure}

In the case where $\delta=3$ we substitute \eqref{kap} for the stationary distribution $\rho(v)$ and therefore, asymptotically in time $t$,
\begin{equation}\label{st-corr}
c(t)\sim t^{1-2\kappa/3}
\end{equation}
(assuming $\kappa > 3/2$). This rough calculation indeed provides a fairly reasonable estimate when $\kappa > 2$, as can be seen in Fig.~\ref{ltt}(a) for a comparison of (\ref{st-corr}) with Monte Carlo results. That is consistent with the discussion in Sec.~\ref{sut}. The long-time tails are entirely due to the active medium and the low friction at high speeds.  Referring again to space plasmas, measuring the time evolution of a specific space plasma parcel is practically very difficult given that the observer (satellite) does not move with the solar wind expansion. Our estimate \eqref{st-corr} offers a specific prediction however.  Long-time tails have been reported for driven granular fluids in e.g.~\cite{zippelius-prl-2009}. As another consequence, by time-integration of $c(t)$, the position is seen to be superdiffusive for $\kappa < 3$ with $\langle (x_t-x_0)^2 \rangle \sim t^{1+f}$ with $f=2-2\kappa/3>0$.  Such behavior is not unseen for tracer particles in bio-active media; see e.g. \cite{caspi-prl-2000}. 

For $1<\delta < 3$ when we substitute in \eqref{eq:ct_int} the expression \eqref{stret} for $\rho(v)$: for large times $t$, we get
\begin{equation}\label{strettime}
c(t) \sim 
 \exp \left[ - \bar{k}\;(\gamma_0\,t)^{\frac{3-\delta}{3}}\right]
 \end{equation}
with $\bar{k} \propto \kappa^{\delta/3}/(3-\delta)$.  
We see that prediction compared with the simulation in Fig.~\ref{ltt}(b) for two values of $\delta=2,\, 2.5$; $c(t)$ for lower values of $\delta$ are more difficult to evaluate.
In the passive case, $A=0$ as for \eqref{maxw}, we have exponential decay in time, reflecting the dilute nature of the thermal bath.  The same happens for $A\neq 0$ (active case) when $\delta=0$ where friction remains prominent (and constant) even at large speeds.  

\begin{figure}[t]
 \centering
 \includegraphics[width=11 cm]{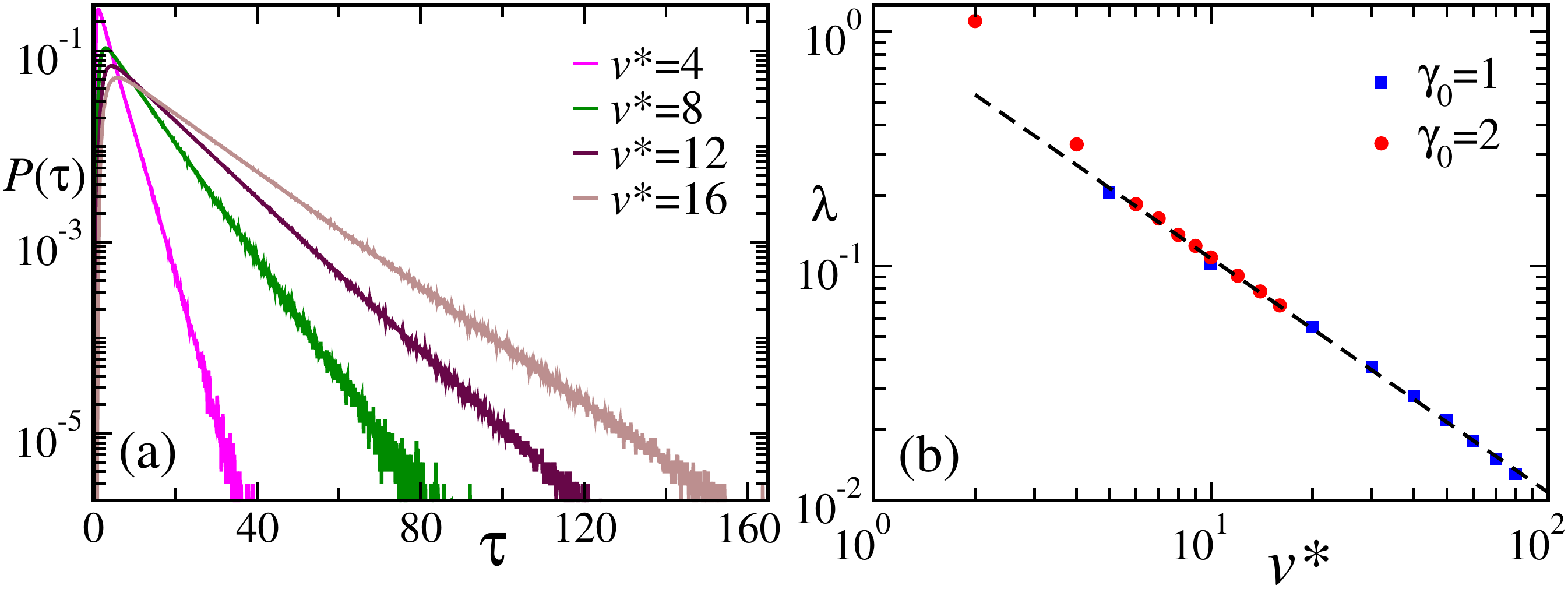}
 \caption{(a) Plot of $P(\tau)$ {\it vs} $\tau$ in semi-log scale for different values of $v^*$ with $\gamma_0=2$ and a fixed width $w=1.$ (b) Plot of the corresponding rate $\lambda$ {\it vs} $v^*$ for two different values of $\gamma_0.$ The dashed line indicates the expected $1/v^*$ behavior. The other parameters are $T=1=L=A=\nu_0$ here. 
 }
 \label{fpp}
\end{figure}

Along similar lines, we also provide an estimate for the 
first passage time probability $P(\tau)$ for the particle  
to remain in a velocity window $[v^* - w\,,\, v^*+w]$ up to a time $\tau.$  For a purely diffusive particle, the first passage time probability in a bounded region decays exponentially with a decay rate proportional to the diffusion constant \cite{Redner}. Using Eqs.~(\ref{intins})-(\ref{clt}), \ie, the effective diffusion picture at low $T$ and large $v^*,$ and translating the result of Ref.~\cite{Redner} to our case, 
 we expect,
\bea
P(\tau) \sim \exp[-\lambda \tau],\;\; \text{with}\; \lambda \propto A^2\,L / (w^2\, v^*) \label{lam-def}
\eea
for large $v^*.$  The average first passage time $\lambda^{-1}$ increases linearly with $v^*$ which is a signature of the trapping in the velocity space discussed before.  Note that in that regime the rate $\lambda$ is independent of the linear friction coefficient $\gamma_0.$
We measure the first passage time probability using numerical simulations to verify this prediction; Fig.~\ref{fpp}(a) shows 
plots of $P(\tau)$ {\it vs} $\tau$ for different values of $v^*$ which clearly shows the exponential decay. The corresponding $\lambda$ are plotted as a function of $v^*$ in Fig.~\ref{fpp}(b) -- the expected $1/v^*$ behavior is seen as $v^*$ increases. \\

%

\section{Conclusions}
 The main result of the paper is that active forces produce suprathermal stationary velocity distributions and long time-tails in the autocorrelation. The {\em activity} of the environment can be so simple as modeled by a tumbling force with a fixed magnitude with a tumbling rate that depends on the speed. The suprathermal
distributions range from power laws over exponentials to Maxwellians, and the time autocorrelation ranges from algebraic to exponential.  The result on long-time tails indicates a persistence in the velocity (or the emergence of extra inertia $\sim \kappa^{-1}$ at high speeds), which in point of fact makes contact with an aspect of self-propelled particles. At the same time it widens the scope of standard activity modeling as for active {\it biological} media, reaching out to and including astrophysical and possibly cosmological processes. Suprathermal behavior and long time-tails carry clear signatures of activity and correspondingly vanish in the absence of the nonconservative force field.

On a more speculative note, apart from space plasmas the importance for equilibration times in cosmological plasmas may be even bigger. In light of the derived long time-tails it indeed cannot be excluded that the usual short-time thermal relaxation assumptions in the derivation of the Kompaneets equation (where photons are treated in contact with electrons having a Maxwellian velocity distribution) cannot be withheld; cf. also~\cite{spaceroar}.

\begin{acknowledgments}
TB acknowledges support from Internal Funds KU Leuven.  CM is grateful to Pierre de Buyl, Bidzina Shergelashvili and to Marian Lazar for inspiring discussions. UB acknowledges support from Science and Engineering Research Board, India under Ramanujan Fellowship (Grant No. SB/S2/RJN-077/2018).
\end{acknowledgments}


\end{document}